\documentclass[11pt,a4paper]{article}

\usepackage{fontspec}
\usepackage{xeCJK}
\xeCJKsetup{CJKspace=true}
\setCJKsansfont[
  BoldFont=UnDotumBold.ttf,
  AutoFakeSlant=0.15
]{UnDotum.ttf}
\setCJKmonofont[
  BoldFont=UnDotumBold.ttf,
  AutoFakeSlant=0.15
]{UnDotum.ttf}
\usepackage[a4paper,margin=2.5cm]{geometry}
\usepackage{authblk}
\usepackage{microtype}
\usepackage{setspace}
\usepackage{caption}
\usepackage{etoolbox}
\AtBeginEnvironment{thebibliography}{\setlength{\itemsep}{0pt}\setlength{\parskip}{0pt}}
\usepackage{pdflscape}
\usepackage{graphicx,amsmath,amssymb,booktabs,array,url,cite}
\usepackage[hidelinks]{hyperref}
\graphicspath{{figures/}}

\title{%
Semantic Communities and Boundary-Spanning Lyrics in K-pop:
A Graph-Based Unsupervised Analysis
}

\author[1]{Oktay Karaku\c{s}\thanks{Corresponding author: \href{mailto:karakuso@cardiff.ac.uk}{karakuso@cardiff.ac.uk}}}
\author[1,2]{Can Eyüpoğlu}
\affil[1]{Department of Computer Science and Informatics, Cardiff University, Cardiff, United Kingdom}
\affil[2]{Department of Computer Engineering, Turkish Air Force Academy, National Defence University, Istanbul, Türkiye}

\date{}

\begin{document}
\maketitle

\begin{abstract}

We examine how lyric representation shapes computational similarity using four 384-dimensional multilingual MiniLM representations of 7,891 K-pop songs by 1,456 artists from charts spanning 2000 to October 2023. Whole-song truncated embeddings provide the strongest standardized separation between neighbouring and random song pairs (1.995), compared with 1.931 for chunked documents, although chunked representations show greater adjusted Rand agreement and line-based representations yield higher modularity. Using whole-song embeddings, we construct a weighted cosine nearest-neighbour graph ($k=30$). A representative Louvain partition identifies 21 operational communities, without a corresponding robust plateau in the Markov Stability analysis. AI-assisted examination of selected lyrics and terms suggests overlapping patterns of memory, separation, longing, self-reflection, attraction, and performance. Community membership is associated with character-script composition (Cram\'er's $V=0.294$; permutation $p=0.001$). High-participation songs exhibit broader cross-community connectivity, while participation has weak negative associations with embedding-relative novelty. These findings show how representation choices and textual characteristics influence the network organization of a cultural corpus, subject to limitations from truncation, graph construction, and the absence of independent human semantic validation.

\end{abstract}

\par

\section{Introduction}

Song lyrics provide a rich record of how popular music expresses
experiences, relationships, identities, and cultural meanings. Their
significance emerges not only from the words they contain, but also
from the imagery, forms of address, repetition, and narrative
progression through which those words are organised. These features
make lyrics particularly interesting for computational analysis:
although individual songs invite close interpretation, large
catalogues offer an opportunity to examine how patterns of expression
recur, overlap, and change across a musical repertoire.
Music-information-retrieval research has established a broad
foundation for the computational study of musical content and
organisation \cite{Downie2003,Schedl2014}, while previous work on
lyrics has explored their usefulness for music classification and
the relationship between textual expression and musical emotion
\cite{FellSporleder2014,MihalceaStrapparava2012}. Yet understanding
how songs relate to one another across an entire catalogue raises
a question that is not resolved by assigning predefined categories:
what aspects of lyrical similarity become visible when the songs
themselves determine the structure of the analysis?

Recent developments in language representation offer one way to
approach this question. Sentence embeddings provide numerical
descriptions of textual content in a shared vector space, making
it possible to compare songs without first assigning them to
particular themes, genres, or emotional categories. Sentence-BERT
and subsequent multilingual knowledge-distillation methods have
provided an important methodological basis for constructing such
representations \cite{ReimersGurevych2019,ReimersGurevych2020}.
Nevertheless, the apparent simplicity of comparing embeddings
conceals a consequential analytical decision: the representation
of a song depends on how its lyrics are presented to the encoder.
A complete lyric treated as a single document does not necessarily
emphasise the same information as a representation assembled from
shorter passages or individual lines. Repeated hooks may contribute
substantially to a line-based representation, whereas removing
duplicate lines changes their influence. Similarly, truncating
a longer lyric may preserve its opening material while excluding
expressions that appear later in the song. Consequently, different
representations of the same catalogue can produce different
relationships between songs, even when the underlying embedding
model remains unchanged.

These differences matter because computational descriptions of a
catalogue are often constructed from the relationships between its
individual items. When songs are connected according to their
embedding similarity, the resulting network provides a way to
examine the organisation of the repertoire without imposing a
thematic classification in advance. Related approaches have
combined document representations with sparsified similarity
networks and multiscale community analysis in other textual
domains, including clinical incident reports
\cite{AltuncuEtAl2019}, while network-based methods have also
contributed to the large-scale study of cultural artefacts
\cite{Schich2014}. Within a lyric network, densely connected
regions may reveal groups of songs that share aspects of their
computational representation. Community-detection methods offer
a means of describing this structure \cite{Newman2006,Blondel2008},
but the resulting communities should not automatically be
interpreted as distinct or naturally occurring semantic categories.
Their composition depends on the representation, the construction
of the network, and the scale at which its organisation is examined.

The question of scale is especially important when interpreting
community structure. A partition that provides a useful summary
at one resolution may conceal meaningful variation within its
communities or exaggerate distinctions between neighbouring
regions of the network. The known resolution limitations of
modularity-based community detection
\cite{FortunatoBarthelemy2007} therefore motivate examining
whether the observed organisation persists under an alternative
view of network structure. Markov Stability provides such a
perspective by studying community organisation across diffusion
times \cite{DelvenneYalirakiBarahona2010}. Rather than treating
agreement between methods as a prerequisite for meaningful
analysis, comparing their behaviour can help distinguish a
convenient operational partition from a more persistent
multiscale structure. This distinction is particularly relevant
to cultural material, where expressions and themes may overlap
rather than form sharply separated groups.

Interpreting the resulting communities introduces a further
challenge. Computational proximity does not, by itself, explain
why particular songs are grouped together. Similarities may
reflect recurring imagery or expressions, but they may also
be influenced by language use, character-script composition,
forms of address, and the repeated phrases characteristic of
popular songwriting. These dimensions need not align neatly:
songs that evoke related experiences may differ considerably
in their linguistic form, while songs sharing a particular
vocabulary may employ it in different expressive contexts.
Community interpretation therefore benefits from examining
both lexical content and observable textual characteristics,
while recognising that neither provides an independent
validation of semantic meaning.

The relationships between communities are equally informative.
Although community detection assigns songs to groups, the
underlying similarity network retains connections that extend
across those assignments. Some songs may be strongly associated
with a relatively local neighbourhood, whereas others connect
to several regions of the catalogue. Such boundary-spanning
patterns offer a complementary perspective on how lyrical
expressions are distributed across the network. They also
raise an interesting distinction between connectivity and
novelty. A song that connects otherwise separated communities
is not necessarily unusual in the embedding space; its
position may instead reflect the combination of familiar
features associated with different groups. Examining
cross-community participation alongside embedding-relative
novelty allows these possibilities to be considered without
assuming that network position directly measures artistic
originality or cultural innovation.

In this study, we investigate these relationships using a
chart-derived corpus of 7,891 K-pop songs spanning 2000 to
October 2023. K-pop provides a relevant setting because its
lyrical repertoire encompasses varied forms of expression,
including multilingual and mixed-script material, within a
recognisable musical and cultural context. Our analysis
proceeds through three connected questions. We first examine
how alternative lyric representations influence the
separation of neighbouring songs from random pairs and the
repeatability of the resulting community partitions. We
then investigate the organisation of a selected similarity
network, characterising its operational communities through
lexical evidence, representative songs, and character-script
composition while assessing their behaviour across
analytical scales. Finally, we consider how songs connect
across community boundaries and whether such connectivity
corresponds to embedding-relative novelty.

Taken together, these analyses seek to clarify how the
computational representation of lyrics shapes the
organisation that can be observed in a large cultural
corpus. The contribution is not a definitive taxonomy
of K-pop themes, nor an independently validated account
of lyrical meaning. Instead, the study connects
representation choices to network geometry, community
interpretation, and boundary-spanning behaviour, providing
an interpretable framework for examining the relationships
among songs while making the methodological conditions
of those interpretations explicit.

\section{Methods}
\label{sec:methods}

\subsection{Corpus and preprocessing}

The study uses a chart-derived collection of K-pop lyrics assembled
from records associated with Melon's Monthly Chart Top~100 rankings.
The source archive contains 25,876 JSON records covering the period
from 2000 to October 2023. Automated parsing and validation initially
identified 7,983 songs by 1,485 artists. A subsequent
catalogue-level consistency assessment excluded 92 songs by 29
artists that were identified as falling outside the intended
K-pop and Korean-market scope. Importantly, the presence of
English-language lyrics or code-switching was not, in itself,
considered grounds for exclusion. The resulting corpus comprises
7,891 songs by 1,456 artists and represents the documented
chart-derived collection within the defined period and market scope.

The historical screening documentation describes a procedure for
identifying possible exclusions among artists represented by at
least two songs. Under this procedure, candidates were required
to have a mean Hangul-character ratio below 0.05, at least
90\% English-dominant songs, and no established identification
with the K-pop or Korean music market. Although these criteria
provide an account of the intended screening logic, the available
records do not establish that their application produced the
final set of 7,891 songs. We therefore report the documented
corpus counts without treating the precise relationship between
the screening procedure and the final exclusions as verified.
Individual songs constitute the observational units throughout
the network analysis, although songs associated with the same
artist, as well as alternate versions of particular recordings,
may introduce dependencies between observations.

The retained data include the source lyrics together with their
associated song and artist identifiers. Preprocessing removes
lines containing only whitespace and strips leading and trailing
whitespace from the remaining lines. No translation, stemming,
or language-specific lemmatization is applied, and repeated
lines, including recurring hooks and refrains, are preserved
in the source text. These decisions retain linguistic and
structural features that may be relevant to lyrical expression,
although their influence varies with the subsequent
representation strategy.

The principal analysis uses an existing set of whole-song
truncated embeddings rather than vectors obtained by pooling
the preprocessed lyric lines. The historical documentation does
not establish the precise truncation limit, tokenizer settings,
truncation direction, or text-normalization operations applied
when those vectors were generated. Because these details cannot
be recovered reliably from the available records, we leave
them explicitly unresolved rather than reconstructing them
from contemporary software defaults. This limitation is
particularly relevant when interpreting differences between
whole-song and line-based representations.

\subsection{Song representations and similarity graph}

To examine how the representation of lyrics influences
computational similarity, we compared four sets of
song-ID-aligned embeddings. Each consists of 384-dimensional
vectors generated using
\texttt{paraphrase-multilingual-MiniLM-L12-v2}, a multilingual
sentence-embedding model grounded in the Sentence-BERT
framework and multilingual knowledge distillation
\cite{ReimersGurevych2019,ReimersGurevych2020}.
The available documentation does not identify the exact
checkpoint revision or complete encoding-time configuration.

The four representations differ in how the textual material
contributes to the final song vector. The whole-song truncated
representation encodes lyrics at the document level, subject
to the encoder's truncation behaviour. The chunked-document
alternative represents lyrics through shorter textual
segments. In the line-mean representation, embeddings of
individual lyric lines are averaged, so repeated lines
contribute repeatedly to the resulting vector. The
unique-line-mean representation instead removes duplicate
lines before aggregation, reducing the influence of repeated
material. Together, these alternatives allow us to examine
whether different treatments of document length, textual
segmentation, and repetition affect the relationships
identified between songs. The precise chunking configuration
and the whole-song encoding settings are not documented.

For each representation, cosine similarity was used to
identify the $k=30$ nearest neighbours of every song.
Neighbour retrieval employed float32 cosine-distance queries
with 31 returned slots per song. The first returned slot
was discarded, and the remaining 30 entries were retained.
Directed neighbour selections were then combined through
an undirected union. Edge weights were obtained as one
minus the returned cosine distance, with a reciprocal
insertion overwriting an existing weight when the same
edge was encountered again.

Although the retrieval procedure was intended to discard
the query song itself, ties in the returned neighbour
ordering could leave self-loops among the retained entries.
These were not subsequently removed. The final whole-song
network contains 14 such loops, and the graph statistics
therefore reflect the network as constructed rather than
an idealized loop-free nearest-neighbour graph.

For a retained edge connecting songs $i$ and $j$, the
corresponding similarity weight is

\begin{equation}
w_{ij}=
\frac{\mathbf{v}_i^\top\mathbf{v}_j}
{\|\mathbf{v}_i\|_2\|\mathbf{v}_j\|_2},
\label{eq:cosine_weight}
\end{equation}

where $\mathbf{v}_i$ and $\mathbf{v}_j$ denote the
respective song embeddings. The selected whole-song graph
contains 7,891 nodes and 209,805 edges, forming a single
connected component. Its mean degree is 53.18, with a
median of 42 and a range from 30 to 539. In accordance
with the graph representation, each retained self-loop
contributes twice to the degree of its associated node.

Representation quality was assessed through the separation
between the cosine similarities of nearest-neighbour pairs
and those of randomly sampled song pairs. Specifically,
we calculated the difference between the two mean
similarities and standardized it using the pooled standard
deviation of their distributions. This statistic describes
the extent to which the representation distinguishes
locally similar songs from a random-pair baseline; it
does not establish the semantic validity of the resulting
neighbourhoods.

The random comparison comprised 236,730 ordered,
non-self song-pair draws for each representation,
corresponding to 30 sampled partners per anchor song,
with replacement. Uncertainty was summarized using
1,000 bootstrap resamples of anchor songs. The resulting
percentile intervals are conditional on the existing
embeddings, retrieved neighbourhoods, and sampled
comparison partners. They should not be interpreted
as accounting for uncertainty arising from regenerating
the embeddings, reconstructing the network, or sampling
a different collection of artists and songs.

We additionally examined the repeatability of community
detection under different initializations. Louvain was
run 20 times for each representation at resolution 2.0,
using seeds 0--19. Agreement between the resulting
partitions was summarized by mean pairwise normalized
mutual information (NMI) and adjusted Rand index (ARI).
These measures indicate how consistently a representation
supports a particular partition under the tested
initializations, rather than whether the communities
correspond to independently established semantic
categories. The complete representation comparison
and repeated-run results are reported in Supplementary
Tables~S1 and S2.

Throughout the analysis, reported modularity refers
to standard weighted modularity evaluated at
$\gamma=1$. This evaluation convention is distinct
from the resolution parameter of 2.0 used to obtain
the Louvain partitions.

\subsection{Operational partition and alternative scale assessment}

The principal community analysis uses the whole-song
similarity graph and a Louvain partition obtained at
resolution 2.0 \cite{Blondel2008,Newman2006}.
Rather than selecting the run with the highest
modularity, we identified a representative partition
by comparing each of the 20 seeded runs with the
remaining 19. Seed 19 achieved the highest mean NMI
against the other partitions and was therefore
retained for subsequent analysis.

For this selected run, the mean NMI relative to the
other partitions is 0.641956 and the corresponding
mean ARI is 0.531412. Its standard weighted
modularity, evaluated at $\gamma=1$, is
$Q=0.420242$. The selected partition contains
21 communities, ranging in size from 38 to
604 songs. Its role is to provide a reproducible
operational grouping for describing the network,
not to establish that the corpus contains exactly
21 naturally distinct lyrical categories.

Because a single community-detection resolution
may obscure alternative forms of organisation,
we also examined the network using continuous-time
Markov Stability
\cite{DelvenneYalirakiBarahona2010}.
The assessment was conducted with PyGenStability
\cite{ArnaudonEtAl2024}, using the full
continuous-time normalized-flow formulation.
We evaluated 13 logarithmically spaced Markov
times between 0.1 and 100, with ten optimization
trials at each time.

Across the evaluated times, the numbers of
communities were 187, 92, 47, 24, 16, 10,
6, 4, 3, 3, 2, 2, and 2, respectively.
To distinguish persistent structure from
short-lived changes in partition resolution,
we defined a descriptive plateau using three
conditions: the same nontrivial community count
had to occur at three or more consecutive
scales; mean normalized variation of information
(NVI) across distinct trial pairs had to be
no greater than 0.15 at each scale; and
adjacent-scale NVI between the best partitions
had to be no greater than 0.15.

No plateau satisfying these criteria was found
within the relevant range of 11--42 communities.
The broader scan identified a robust two-community
plateau only toward the coarser end of the tested
time grid. This observation does not imply that
meaningful organisation is absent at intermediate
scales. Rather, it indicates that the available
analysis did not establish a comparably persistent
partition within the range most relevant to the
operational Louvain grouping.

For a more direct comparison, we selected the
24-community Markov Stability partition at
time 0.562341 on the basis of its within-scan
stability diagnostics. This choice was made
independently of its agreement with Louvain.
Compared with the 20 Louvain partitions, the
selected multiscale partition achieved a mean
NMI of 0.660 (range 0.624--0.713) and a mean
ARI of 0.550 (range 0.495--0.625).
These values indicate substantial overlap in
the broad organisation identified by the two
approaches, alongside differences in their
individual community assignments.

The multiscale results consequently provide
context for interpreting the selected
21-community partition rather than independent
confirmation of its precise boundaries.
The assessment is also constrained by the
finite Markov-time grid and number of
optimization trials. Available software
documentation identifies Python 3.10.19
and PyGenStability 0.2.5, although it does
not verify that this was the environment
used to generate the reported scan.
The available scripts are consistent with
the documented multiscale procedure but
cover only that component of the broader
analytical pipeline.

We therefore retain the representative
Louvain partition as a practical analytical
summary while acknowledging the
resolution-dependent nature of community
detection \cite{FortunatoBarthelemy2007}.

\subsection{Script and community characterization}

To investigate whether the communities differ
in their textual composition, we quantified
character-script usage directly from the
retained lyrics. Script shares were calculated
among Unicode alphabetic characters, with
characters classified by their Unicode names
as Hangul, Latin, or other. Whitespace,
punctuation, and numerical characters were
excluded from the denominator, while
repeated lyric passages remained part
of the calculation.

Songs were assigned to descriptive
script-composition strata using a hierarchical
set of thresholds. A song was classified
as Hangul-dominant when its Hangul share
was at least 0.80. Among the remaining
songs, those with a Latin share of at
least 0.80 were classified as
Latin-dominant. Songs not meeting either
criterion were classified as mixed
Hangul--Latin when the combined share
of these two scripts was at least 0.80.
All remaining songs were assigned
to an other-script category. A separate
no-letters category was defined for
records containing no alphabetic characters,
although no such records occurred
in the corpus.

These categories describe the visible
writing systems present in the lyrics.
They are not language-identification
outputs and should not be interpreted
as validated distinctions between
Korean-language, English-language,
and multilingual songs.

The association between script composition
and community membership was assessed
using a Pearson contingency statistic.
Statistical significance was evaluated
through 999 global permutations of song
labels, implemented using NumPy's
default random-number generator with
seed 42. The permutation $p$-value
was calculated using the plus-one
estimator. We emphasize this result
because some cells in the contingency
table have small expected counts.
Nevertheless, the permutation procedure
assumes exchangeability at the song
level and does not account for
dependencies arising from shared
artists, alternate versions, or
connections within the similarity
network.

Community interpretation combined
quantitative profiles with a bounded
qualitative examination of their
lyrical content. For each community,
we considered its overall profile,
20 ranked discriminative terms derived
from within-community versus
outside-community TF--IDF evidence,
and selected representative songs.
The purpose was to identify recurring
expressive characteristics that could
help explain the observed computational
groupings, rather than to assign
exhaustive thematic classifications
to every lyric.

Where sufficient records were available,
the review considered 12 unique songs
per community. Candidate songs were
selected by alternating between
centroid-based rankings and
within-community weighted PageRank
rankings. The examination included
songs at ranks 1, 4, 8, and 12,
with additional checks at rank 6
where these were informative.
This procedure combined proximity
to a community's embedding centre
with prominence in its internal
network structure, although neither
criterion guarantees thematic
representativeness.

The interpretive synthesis was
AI-assisted and exploratory.
The documented review did not
involve an external model or API
call, and it did not employ a
formal motif-coding protocol or
independent human semantic
validation. Community labels
therefore describe patterns
suggested by the bounded evidence.
They are not estimates of thematic
prevalence or community purity,
and they do not imply that every
song assigned to a community
shares the same interpretation.

Where qualitative confidence
is reported, it refers to the
strength and coherence of the
available descriptive evidence,
not to a statistical confidence
interval. Community profiles,
discriminative terms, script
distributions, and the complete
21-community interpretation table
are provided in Supplementary
Table~S7. Additional script counts
and composition summaries appear
in Supplementary Table~S8 and
Supplementary Figure~S2.

\subsection{AI-assisted manuscript preparation}

AI-assisted tools were used to support exploratory interpretation of community characteristics and to improve manuscript language, clarity, and organisation. The authors reviewed the outputs and take full responsibility for the analyses, interpretations, and final manuscript.

\subsection{Boundary participation and novelty}

The community partition provides
a description of densely connected
regions of the similarity graph,
but it does not capture the full
distribution of a song's connections
across the network. To examine
relationships extending beyond
individual communities, we
characterised each song using
weighted participation, external
connectivity, and two
embedding-relative measures
of novelty.

Let $s_{ic}$ denote the total
weight of edges connecting song
$i$ to songs assigned to
community $c$. The total
strength of song $i$ is
$s_i=\sum_c s_{ic}$.
Its weighted participation
coefficient is defined as

\begin{equation}
P_i=1-\sum_{c=1}^{C}
\left(\frac{s_{ic}}{s_i}\right)^2 .
\label{eq:participation}
\end{equation}

Participation increases as a
song's incident edge weight
becomes more evenly distributed
across different communities.
It therefore describes the
breadth of a song's weighted
community connections rather
than simply whether the song
has links outside its own group.

To distinguish these properties,
we also calculated the fraction
of total strength directed
toward other communities:

\begin{equation}
F_i^{\mathrm{ext}}=
\frac{\sum_{c\ne c_i}s_{ic}}{s_i},
\label{eq:external_fraction}
\end{equation}

where $c_i$ denotes the
community to which song $i$
is assigned. Unlike
participation, this measure
focuses specifically on
the proportion of connectivity
lying outside the assigned
community.

We operationally defined
boundary songs as those
whose participation was
at or above the empirical
90th percentile. In the
selected graph, this
threshold was 0.832139,
identifying 790 boundary
songs and 7,101 remaining
songs. The definition is
relative to the observed
participation distribution
and the selected community
partition; it is not a
universal criterion for
boundary-spanning behaviour.

For additional descriptions
of how connectivity is
distributed, we defined
each song's community-strength
share as

\begin{equation}
p_{ic}=\frac{s_{ic}}{s_i}.
\label{eq:community_share}
\end{equation}

Using these shares,
normalized community
entropy is

\begin{equation}
H_i=-\frac{
\sum_{c:p_{ic}>0}
p_{ic}\log p_{ic}
}{\log C}.
\label{eq:entropy_reach}
\end{equation}

For the selected partition,
$C=21$, giving a
normalization factor
of $\log(21)$.
Entropy captures the
evenness of a song's
weighted connections
across communities,
while participation
summarizes a related
aspect of their
concentration.

We also measured the
number of distinct
external communities
reached by each song:

\begin{equation}
R_i=
\left|\{c:c\ne c_i,\ s_{ic}>0\}\right|.
\label{eq:external_reach}
\end{equation}

This external-reach
measure counts communities
with positive connection
strength, irrespective
of the magnitude of
those connections.
Participation, external
fraction, entropy, and
external reach are
therefore complementary
descriptions of the
same underlying network
relationships, not
independent sources
of validation.

To examine whether
boundary-spanning
connectivity coincides
with unusual embedding
positions, we considered
two measures of
representation-relative
novelty.

The first describes a
song's distance from
the nearest community
centroid. Let
$\boldsymbol{\mu}_c$
denote the centroid
associated with
community $c$.
Centroid-relative
novelty is defined as

\begin{equation}
N_i^{\mathrm{centroid}}=
1-\max_{c}
\cos(\mathbf{v}_i,\boldsymbol{\mu}_c).
\label{eq:centroid_novelty}
\end{equation}

A higher value indicates
that the song is less
similar to its closest
community centroid
under the selected
embedding representation.
The historical documentation
does not establish the
normalization order used
when constructing the
centroids, and this
detail remains unresolved.

The second measure
characterises novelty
relative to a song's
immediate graph
neighbourhood. It is
defined as one minus
the arithmetic mean
weight of the song's
incident edges:

\begin{equation}
N_i^{\mathrm{local}}=
1-\frac{1}{|\mathcal{E}_i|}
\sum_{(i,j)\in\mathcal{E}_i}w_{ij},
\label{eq:local_novelty}
\end{equation}

where $\mathcal{E}_i$
denotes the stored set
of incident edges.
Any retained self-loop
is included in this
calculation.

These two novelty
measures address
different aspects
of embedding geometry.
Centroid-relative
novelty describes
separation from
community centres,
whereas local novelty
describes dissimilarity
to directly connected
neighbours. Neither
provides an independent
human assessment
of lyrical originality,
semantic innovation,
or cultural novelty.

We additionally
summarized boundary
participation at
the artist level,
using exact artist
strings and retaining
the existing inclusion
threshold of at least
five songs. Artist
aliases, individual
members, solo releases,
and collaborations
were not consolidated.
Consequently, these
summaries describe
the available catalogue
under its recorded
artist identifiers
and should not be
interpreted as
population-level
comparisons of artists.

Artist means were
accompanied by
1,000 song-level
bootstrap percentile
intervals generated
with seed 42.
Because songs were
resampled individually,
these intervals do
not account for
dependencies associated
with shared artists,
catalogue construction,
or graph connectivity.

Finally, differences
between the q90
boundary group and
the remaining songs
were summarized
using two-sided
Mann--Whitney $U$
tests, Cliff's
$\delta$, and
song-level percentile
bootstrap intervals
for mean differences.
These statistics
provide descriptive
comparisons conditional
on the existing
whole-song graph,
community assignments,
and boundary definition.
They do not establish
causal relationships,
and the resampling
procedure does not
preserve dependence
among connected songs,
artists, or alternate
versions.

\section{Representation selection and graph structure}
\label{sec:representation_results}

The choice of lyric representation had a substantial influence on the
similarity relationships recovered from the corpus. Although all four
representations were evaluated using the same $k=30$ nearest-neighbour
construction, they produced markedly different distributions of cosine
similarity. This variation was evident not only among neighbouring
songs but also among randomly paired songs, suggesting that the
representations differed in the extent to which they distinguished
local relationships from the broader similarity structure of the
catalogue.

The two line-based representations produced particularly high
absolute similarities. For the line-mean representation, the mean
cosine similarity among neighbouring songs was 0.9784 (sample
SD = 0.0337), compared with 0.9395 (SD = 0.0722) among randomly
paired songs. The unique-line-mean representation showed a similar
pattern, with corresponding values of 0.9844 (SD = 0.0276) and
0.9550 (SD = 0.0648). In contrast, the chunked-document
representation produced mean similarities of 0.8393 (SD = 0.0724)
for neighbouring songs and 0.6374 (SD = 0.1289) for random
pairs. The whole-song truncated representation yielded the
lowest absolute similarities, at 0.7836 (SD = 0.0894) and
0.5420 (SD = 0.1462), respectively.

These results illustrate why high similarity between nearest
neighbours cannot, by itself, establish that a representation
provides a more informative description of the corpus. The
line-based approaches placed neighbouring songs very close
together in the embedding space, but they also assigned
comparatively high similarity to songs paired at random.
Consequently, the distinction between local neighbourhoods
and the wider catalogue was considerably smaller than the
absolute neighbour similarities might initially suggest.

This difference becomes clearer when the separation between
neighbouring and random pairs is examined directly. The
neighbour-minus-random mean similarity gap was 0.03885
(95\% anchor-bootstrap interval [0.03838, 0.03931]) for
line mean and 0.02945 [0.02900, 0.02994] for unique-line
mean. The corresponding gaps were substantially larger for
chunked-document, at 0.20193 [0.20122, 0.20262], and
whole-song truncated, at 0.24162 [0.24087, 0.24240].
Standardising these differences by the pooled standard
deviation produced the same broad ordering: 0.690
[0.668, 0.711] for line mean, 0.592 [0.570, 0.614]
for unique-line mean, 1.931 [1.914, 1.948] for
chunked-document, and 1.995 [1.978, 2.011] for
whole-song truncated.

The whole-song representation therefore achieved the
strongest observed separation between local neighbours
and the random-pair baseline, although its advantage
over chunked-document was relatively modest on the
standardised measure. The reported intervals describe
variation under the conditional anchor-bootstrap
procedure specified in Methods. They do not constitute
formal pairwise significance tests between
representations, nor do they establish that the
separated neighbourhoods correspond to independently
validated lyrical meanings.

A different pattern emerged when the representations
were evaluated through the community structure of
their corresponding similarity graphs. Across
20 Louvain runs, mean standard weighted modularity
was 0.46498 (SD = 0.00669) for line mean,
0.46872 (SD = 0.00440) for unique-line mean,
0.44927 (SD = 0.00506) for chunked-document,
and 0.41523 (SD = 0.00338) for whole-song
truncated. The line-based representations thus
produced higher modularity than either
document-level alternative, despite their
substantially weaker separation between
neighbouring and random song pairs.

Agreement between repeated community-detection
runs introduced a further distinction. Mean
pairwise NMI and ARI were 0.5779 and 0.4326
for line mean, 0.6252 and 0.4906 for
unique-line mean, 0.6195 and 0.5310 for
chunked-document, and 0.6226 and 0.5027
for whole-song truncated. Unique-line mean
achieved the highest NMI, whereas
chunked-document achieved the highest ARI.
Whole-song truncated remained close to the
best NMI result, although it did not lead
on either partition-agreement measure.

Taken together, these comparisons reveal
that the evaluation criteria capture
different properties of the resulting
representations. Neighbour-versus-random
separation describes the contrast between
local similarity and a corpus-wide baseline,
while NMI and ARI describe the repeatability
of community assignments across
optimisation runs. Modularity, meanwhile,
characterises the concentration of
weighted connections within a particular
partition. Because the four representations
produce different graph topologies,
differences in modularity should be
interpreted descriptively rather than
as direct evidence that one representation
captures lyrical meaning more accurately.

We selected the whole-song truncated
representation for the subsequent
analysis because it provided the
largest observed separation between
neighbouring and randomly paired songs,
while maintaining partition agreement
close to the highest NMI obtained.
This decision reflects a preference
for distinguishing local relationships
from the wider catalogue without
sacrificing substantial repeatability
in the resulting community structure.
It does not imply that whole-song
truncation is preferable under every
criterion. In particular,
chunked-document achieved a similar
standardised separation and higher
ARI and modularity, while
unique-line mean yielded slightly
higher NMI. The selected
representation also retains the
important limitation that truncation
may exclude material appearing
later in a lyric.

\begin{figure}[t]
\centering
\includegraphics[
    width=\linewidth,
    height=0.5\textheight,
    keepaspectratio
]{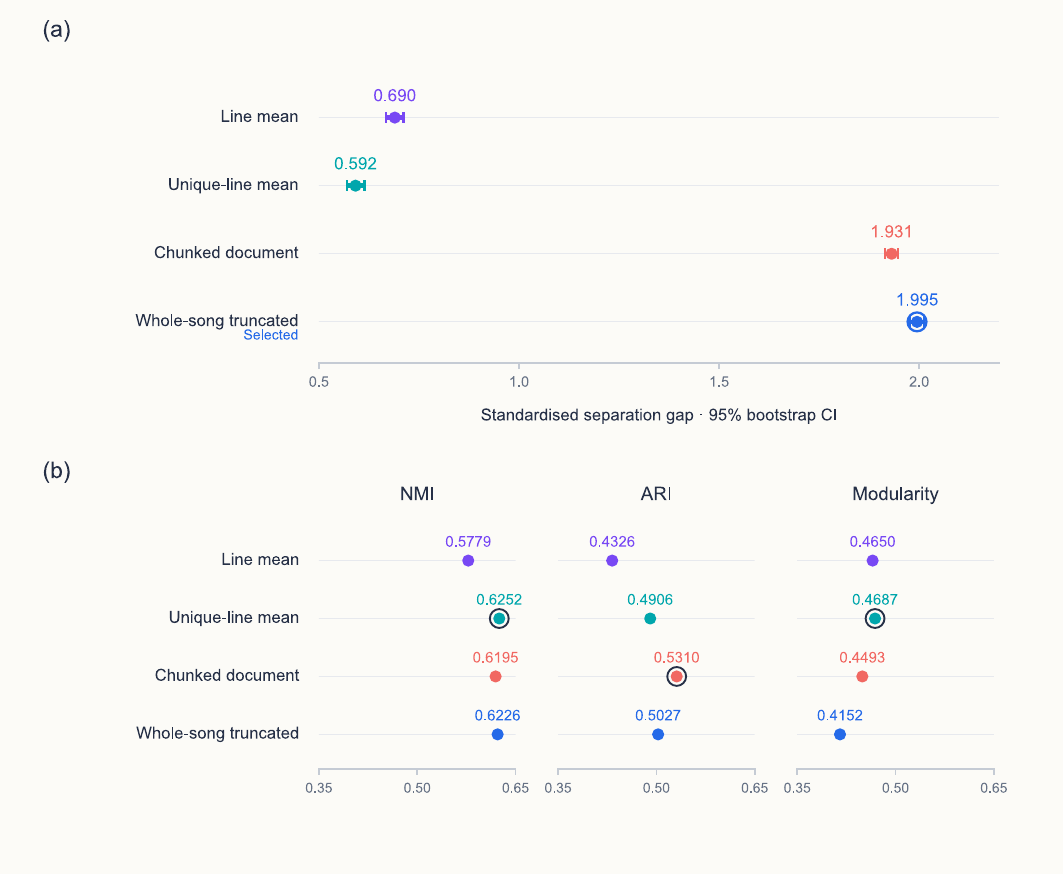}
\caption{Representation comparison under the matched $k=30$ graph
construction for four strategies: line mean, unique-line mean,
chunked document, and whole-song truncated. The figure shows the
standardized neighbour-versus-random cosine separation with its
conditional 95\% anchor-bootstrap interval, together with mean
pairwise NMI and ARI between Louvain runs and standard weighted
modularity. Whole-song truncated is the selected representation;
this selection does not imply superiority on every metric.
NMI and ARI are run-agreement means without uncertainty
intervals, and the separation intervals are not formal
pairwise significance tests.}
\label{fig:representation_geometry}
\end{figure}

Figure~\ref{fig:representation_geometry}
and Table~\ref{tab:main_representations}
summarise these comparisons and
the trade-offs underlying the
representation choice.

\begin{table}[t]
\centering
\caption{Comparison of the four lyric-embedding representations under matched settings (Supplementary Table~S1). Separation gap is the pooled-SD standardized difference between mean neighbour cosine similarity and mean random-pair cosine similarity, with a 95\% anchor-song bootstrap interval (1,000 resamples) that is conditional on the song embeddings and sampled neighbours. Modularity is the mean (SD) across Louvain runs; NMI and ARI are means over 190 pairwise comparisons between runs and are not uncertainty intervals. Whole-song truncated, the selected representation, has the largest separation gap but not the highest NMI, ARI or modularity.}
\label{tab:main_representations}
\small
\setlength{\tabcolsep}{5pt}
\renewcommand{\arraystretch}{1.1}
\begin{tabular}{@{}l c c c c@{}}
\toprule
Representation & Separation gap [95\% CI] & NMI & ARI & Modularity (SD) \\
\midrule
Line mean & 0.690 [0.668, 0.711] & 0.5779 & 0.4326 & 0.46498 (0.00669) \\
Unique-line mean & 0.592 [0.570, 0.614] & 0.6252 & 0.4906 & 0.46872 (0.00440) \\
Chunked document & 1.931 [1.914, 1.948] & 0.6195 & 0.5310 & 0.44927 (0.00506) \\
Whole-song truncated (selected) & 1.995 [1.978, 2.011] & 0.6226 & 0.5027 & 0.41523 (0.00338) \\
\bottomrule
\end{tabular}
\end{table}

Having selected the whole-song representation, we
examined the organisation of its weighted similarity
network. The graph contains 7,891 songs connected
by 209,805 edges and forms a single connected
component. Its mean degree is 53.18, with a
median of 42 and a range from 30 to 539.
Although the underlying construction used
30 nearest-neighbour selections per song,
the undirected union allows individual
nodes to acquire additional connections
when they are selected as neighbours
by other songs. The graph also retains
14 self-loops arising from tied
neighbour similarities, as described
in Methods.

The representative Louvain partition,
obtained at resolution 2.0 using
seed 19, identifies 21 operational
communities. These range in size
from 38 to 604 songs, with a median
community size of 372. The partition
has standard weighted modularity
$Q=0.420242$ when evaluated at
$\gamma=1$. Seed 19 was selected
because it achieved the highest
mean NMI relative to the other
19 runs, rather than because
it maximised modularity.
Its mean agreement with the
remaining runs was 0.641956
for NMI and 0.531412 for ARI.

The resulting network provides
a structured view of relationships
among songs in the selected
embedding space. Its communities
identify regions in which
weighted connections are
relatively concentrated,
offering a practical basis
for examining differences
in lyrical content and
textual characteristics.
Nevertheless, the presence
of 21 communities should
not be understood as evidence
that the repertoire contains
exactly 21 distinct semantic
categories. The partition
is a resolution-dependent
description of the network,
and its boundaries remain
conditional on the selected
representation and graph
construction.

\begin{figure}[t]
\centering
\includegraphics[
    width=\linewidth,
    height=0.6\textheight,
    keepaspectratio
]{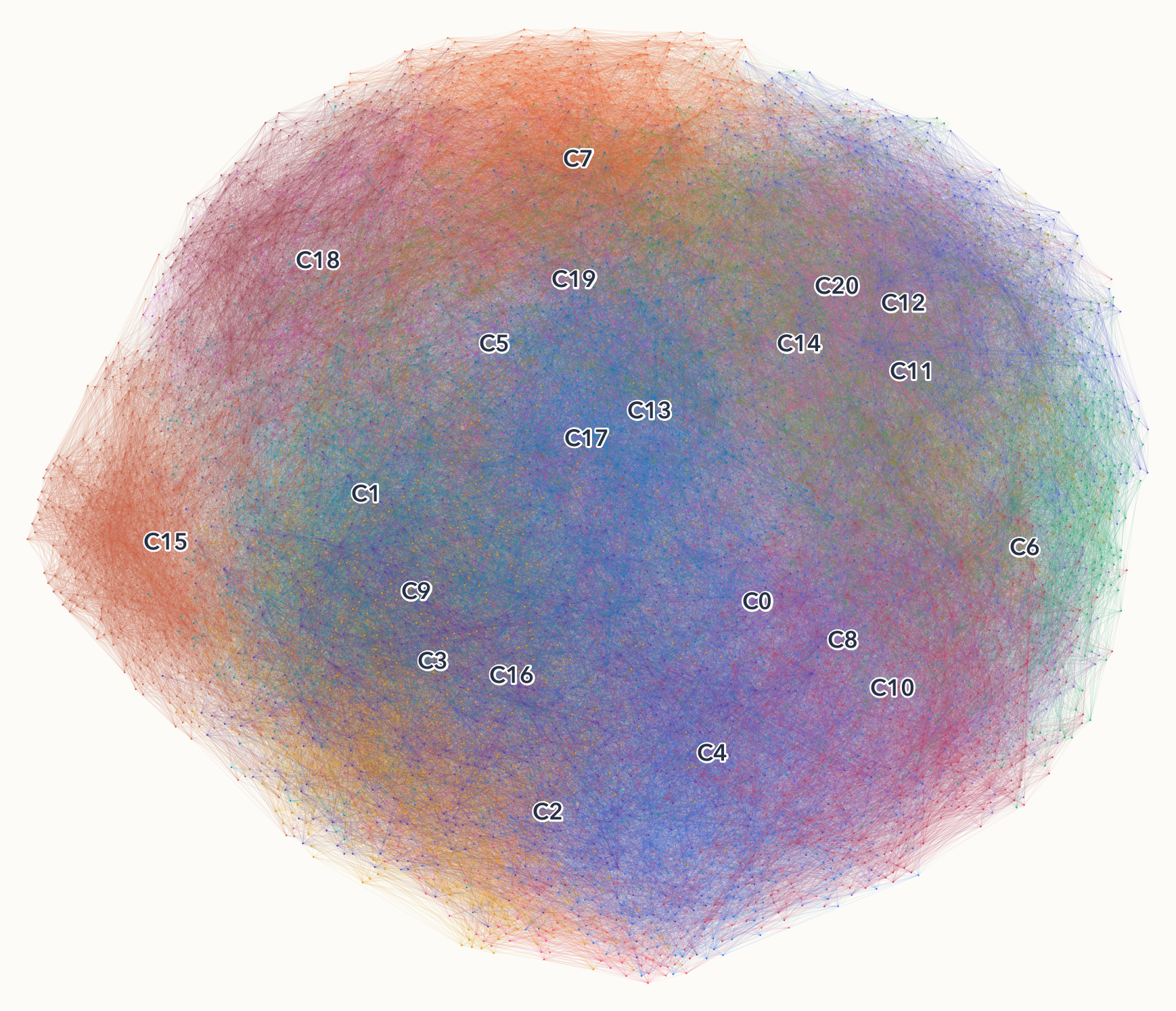}
\caption{Semantic network atlas of the 7,891 songs (nodes) in the
weighted cosine $k$-nearest-neighbour graph ($k=30$; 209,805
edges). Node and within-community edge colours identify the
21 operational Louvain communities of the seed-19 partition
(C0--C20); grey edges connect different communities.
Positions come from a force-directed layout and are
diagrammatic, not quantitative semantic distances;
the layout does not independently validate
community separation.}
\label{fig:semantic_network_atlas}
\end{figure}

Figure~\ref{fig:semantic_network_atlas}
illustrates this operational
organisation. The force-directed
layout makes the community
structure visually accessible,
but the distances between
plotted nodes should not be
interpreted as quantitative
semantic distances or as
independent evidence of
community separation.

To assess whether this organisation persisted
across alternative scales, we compared the
operational Louvain partition with the
continuous-time Markov Stability analysis.
As Markov time increased from 0.1 to 100,
the number of detected communities
progressively decreased, following
the sequence
187--92--47--24--16--10--6--4--3--3--2--2--2.
This progression reflects the transition
from finer subdivisions of the network
towards increasingly coarse descriptions
of its structure.

Importantly, no partition within
the range of 11--42 communities
satisfied the plateau criteria
specified in Methods. A stable
two-community configuration
appeared only towards the
coarsest end of the evaluated
Markov-time range. Thus,
although the analysis reveals
organisation at several
resolutions, it does not
identify a persistent
intermediate-scale partition
that independently establishes
the 21-community Louvain
solution as a uniquely
preferred grouping.

The selected 24-community
Markov Stability partition
nevertheless showed
substantial correspondence
with the Louvain results.
Across the 20 Louvain
runs, its mean NMI was
0.660 (range 0.624--0.713),
while its mean ARI was
0.550 (range 0.495--0.625).
These values indicate that
the two approaches recover
overlapping aspects of the
network's organisation,
even though they do not
produce identical
community assignments.

The comparison therefore
supports a qualified
interpretation of the
selected partition.
The Louvain communities
provide a reproducible
and analytically useful
organisation of the
song network, while
the multiscale analysis
shows that their precise
number and boundaries
should not be treated
as intrinsic properties
of the corpus. The
Markov-time trajectory
and its associated
diagnostics are reported
in Supplementary
Figure~S1 and
Tables~S3--S6.

\section{Community content and script association}
\label{sec:community_results}

The communities identified in the selected similarity network differ
not only in their patterns of connectivity but also in the linguistic
and expressive characteristics of their constituent songs. To examine
these differences, we considered character-script composition alongside
the lexical signatures and representative lyrics associated with each
community. This combination provides a way to investigate how textual
form and lyrical content contribute to the observed organisation,
without assuming that computationally defined communities correspond
to discrete semantic categories.

\begin{figure}[t]
\centering
\includegraphics[
    height=0.55\textheight,
    keepaspectratio
]{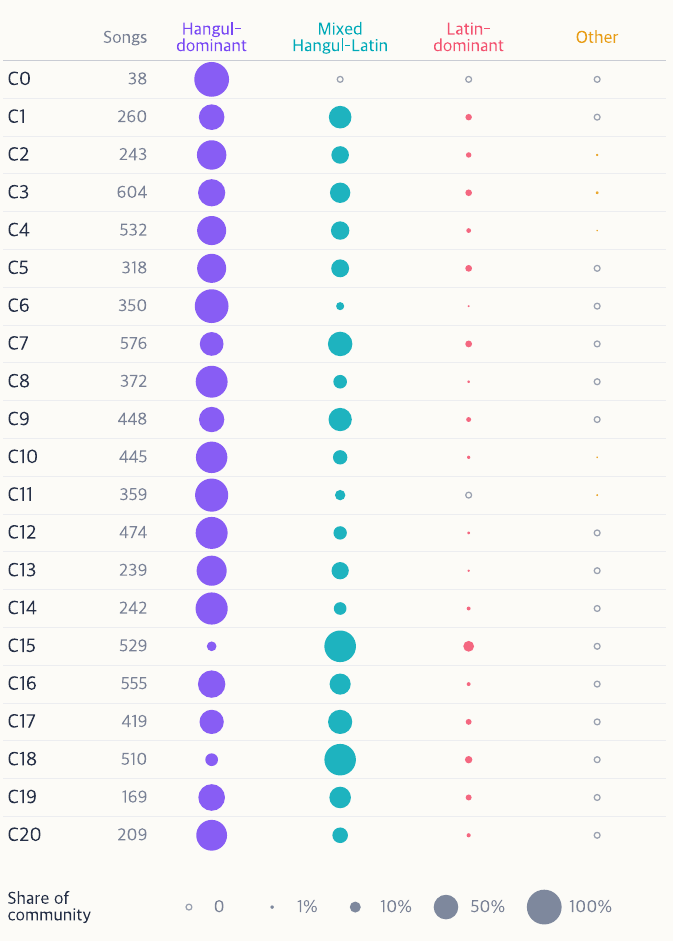}
\caption{Script composition of the 21 operational communities (C0--C20).
Each row shows the within-community proportion of songs in four
character-script categories (Hangul-dominant, mixed Hangul--Latin,
Latin-dominant, other); circle area is proportional to that share on one common
scale, and hollow circles indicate zero songs. The left column gives community
song counts. Script categories are character-composition proxies, not validated
language identification. The association between script composition and
community membership is descriptive ($\chi^2(60)=2048.54$; Cram\'er's
$V=0.294$, bias-corrected $V=0.290$; permutation $p=0.001$) and does not imply
causation.}
\label{fig:script_landscape}
\end{figure}

Character-script composition varies considerably across the
21 operational communities (Figure~\ref{fig:script_landscape}).
Across the full corpus, 4,865 songs were classified as
Hangul-dominant, 2,813 as mixed Hangul--Latin, 205 as
Latin-dominant, and 8 as other-script. The association between
script composition and community membership was substantial,
with a Pearson contingency statistic of
$\chi^2(60)=2048.54$ and Cram\'er's $V=0.294$
(bias-corrected $V=0.290$). The permutation analysis
returned $p=0.001$ from 999 permutations. Because
23 expected cells contained fewer than five observations,
the asymptotic chi-square approximation was not treated
as the primary basis for inference. The permutation result
provides the principal descriptive assessment, subject to
the song-level exchangeability assumptions discussed in
Methods.

These findings indicate that the distribution of writing
systems is not uniform across the selected communities.
However, character-script composition is only one aspect
of the textual variation represented in the network.
The categories describe the relative presence of Hangul,
Latin, and other alphabetic characters rather than
independently identified languages, and they cannot
establish whether a community is organised primarily
around linguistic form, lyrical meaning, or some
combination of the two. Interpreting the association
therefore requires closer examination of the content
and lexical characteristics of individual communities.

\begin{figure}[p]
\centering
\includegraphics[
    width=\linewidth,
    height=0.8\textheight,
    keepaspectratio
]{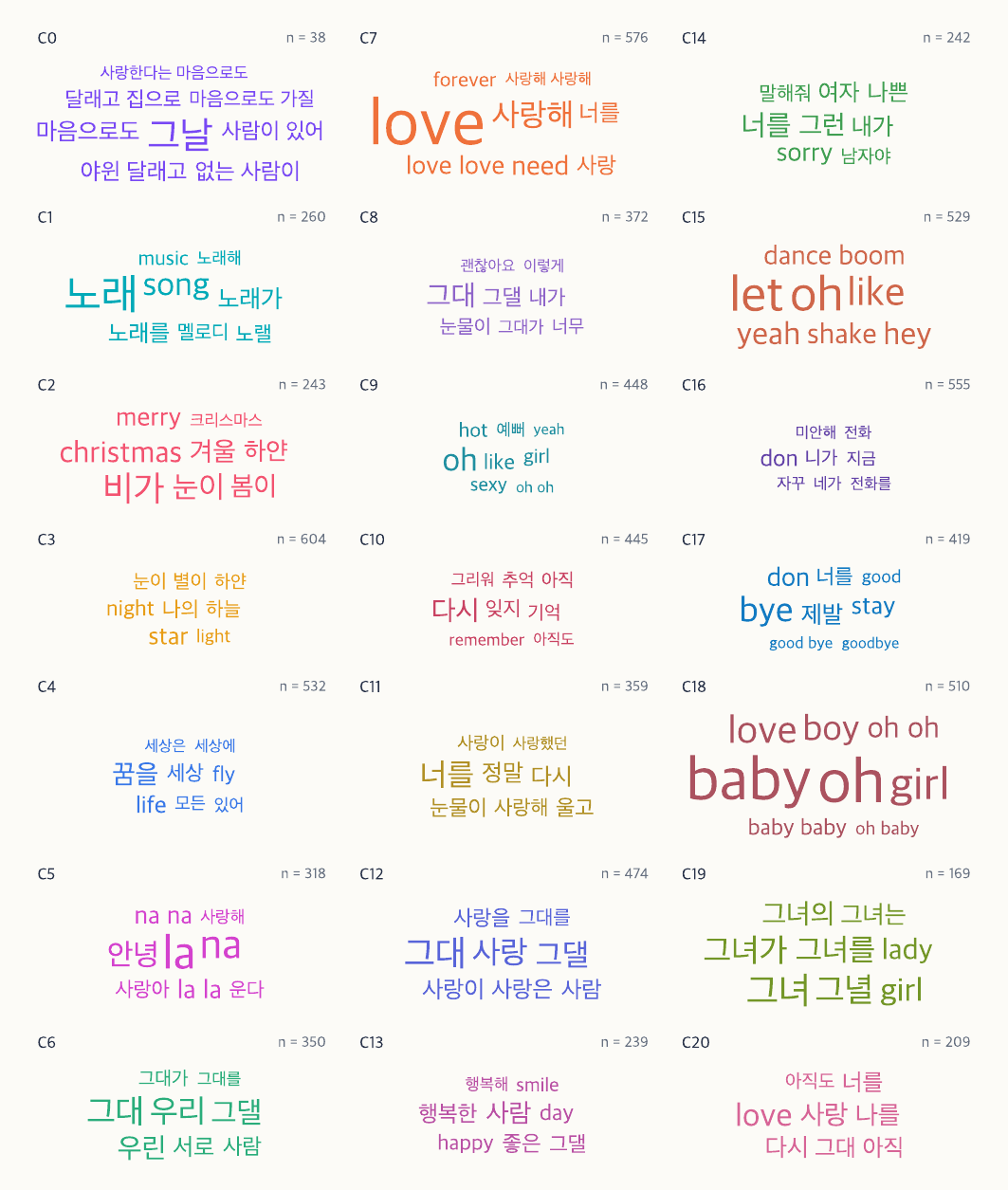}
\caption{Exploratory lexical signatures of the 21 operational communities
(C0--C20), with community song counts. Each panel shows eight terms selected
from the positive TF-IDF differences for that community. Word
bounding-box area encodes the difference on one common global visual
scale across all panels. Word positions carry no semantic meaning. These
signatures are exploratory descriptions of the same embedding graph and
partition, not validated semantic themes.}
\label{fig:community_constellations}
\end{figure}

The community-level lexical profiles reveal several recurring
patterns of expression, although their interpretation differs
across groups. Figure~\ref{fig:community_constellations}
presents the discriminative lexical signatures of all
21 communities, while Table~\ref{tab:main_communities}
summarises selected examples that illustrate the range
of observed characteristics. Rather than treating these
signatures as definitive thematic labels, we use them
together with representative lyrics to examine how
particular forms of expression become concentrated
within different regions of the network.

Community C6 provides an example in which relationship-oriented
content accompanies a relatively homogeneous script profile.
Of its 350 songs, 94.3\% are Hangul-dominant. Its
discriminative vocabulary includes reciprocal and
relational expressions such as \emph{geudae} (you),
\emph{uri} (we/us), and \emph{seoro} (each other).
The reviewed lyrics suggest recurring concerns with
devotion, interpersonal attachment, and changing
relationships. The prominence of reciprocal address
is consistent with these interpretations, although
the vocabulary does not imply that every song
expresses the same emotional situation. C6
therefore illustrates how a recognisable expressive
tendency can coexist with a strongly
Hangul-dominant textual composition.

Other communities exhibit more distinctive
combinations of imagery and narrative context.
C2, which contains 243 songs, is characterised
by vocabulary associated with seasons and weather,
including Christmas, winter, snow, spring,
rain, and related expressions. Representative
songs such as \emph{광화문에서} and
\emph{비가 오는 날엔} place experiences
of separation and remembrance within
seasonal or rainy settings. The recurrence
of this imagery suggests a shared expressive
context rather than a narrowly defined
Christmas-related theme. Seasonal references
appear to provide a setting through which
different relationship narratives are
developed, connecting songs that need
not describe identical experiences.

A related but distinguishable pattern
appears in C8, which contains 372 songs.
The reviewed examples, including
\emph{또 한번 사랑은 가고},
\emph{부디..}, and
\emph{신촌을 못가}, associate
loss and emotional distress with
memories of particular relationships
and places. Here, the interpretive
emphasis falls less on recurring
seasonal imagery and more on the
relationship between separation,
grief, and remembered locations.
Although these concerns overlap
with those identified elsewhere
in the network, their combination
provides a useful description
of the material examined in C8.

Taken together, C2 and C8 illustrate
an important feature of the
community structure. Broad emotional
concerns such as longing, separation,
and memory are not confined to
individual communities. Instead,
different groups may express
related experiences through
particular combinations of
imagery, narrative setting,
and forms of address. The
resulting distinctions are
therefore better understood
as differences in recurring
patterns of expression than
as boundaries between
mutually exclusive themes.

The interaction between lyrical content
and textual form becomes particularly
apparent in C15. This community
contains 529 songs, approximately
82.6\% of which fall within
the mixed Hangul--Latin category.
Its mean Latin-character share
is 0.534, slightly exceeding
the corresponding Hangul share
of 0.466. Representative
titles such as \emph{SHAKE THAT BRASS},
\emph{Dream Girls}, and
\emph{Busan Vacation}, together
with lexical expressions
associated with dancing,
parties, repeated calls,
and audience address,
suggest an emphasis on
performance-oriented
lyrical forms.

The qualitative review
assigned high confidence
to a primarily
language/form-driven
description of C15.
Nevertheless, this
characterisation should
not be interpreted as
a separation between
form and meaning.
References to movement,
celebration, and direct
audience engagement
also convey substantive
content, while repetition
and code-switching may
contribute to the
recognisable structure
of performance-oriented
lyrics. C15 thus
illustrates how
linguistic composition,
lyrical function,
and recurring expressive
content may reinforce
one another within
a computationally
defined community.

Not all communities,
however, support
equally coherent
interpretations.
The bounded
AI-assisted synthesis
classified eight
communities as
theme-driven, ten
as combining content
with language or
formal characteristics,
one as primarily
language/form-driven,
and two as
heterogeneous or
unclear. These
classifications
summarise the
available qualitative
evidence rather
than independently
validated categories.
They do not measure
the proportion of
songs that conform
to a proposed theme,
nor do they establish
that the themes are
exclusive to their
assigned communities.

Two cases illustrate
the importance of
this distinction.
C0 contains only
38 songs, and
several prominent
records are versions
of \emph{가질 수 없는 너}.
Their shared lyrical
material may
contribute to the
apparent coherence
of the community,
making it difficult
to distinguish
a broader expressive
pattern from the
influence of
closely related
recordings.
By contrast,
C16 contains
555 songs and
exhibits familiar
relationship
vocabulary, but
its reviewed
examples describe
a range of
different situations.
Although relational
language is
prominent, the
available evidence
does not support
a single
high-confidence
thematic
interpretation.

These contrasting
examples highlight
why community
coherence cannot
be inferred solely
from the presence
of distinctive
terms or a
concentrated
network structure.
A small community
may appear
consistent because
it contains
multiple versions
of related
material, whereas
a larger community
may bring together
songs that share
vocabulary without
expressing the
same narrative
or emotional
perspective.
The qualitative
evidence must
therefore be
considered in
relation to
community size,
the composition
of the reviewed
sample, and
the possible
influence of
repeated or
closely related
lyrics.

Overall, the
community profiles
suggest that the
selected network
organises songs
according to
overlapping
combinations of
lyrical content,
character-script
composition, and
expressive form.
Some communities
exhibit relatively
recognisable
thematic tendencies,
while others
appear to reflect
mixed linguistic
and semantic
characteristics
or remain
difficult to
characterise
coherently.
The recurrence
of motifs such
as love, longing,
separation,
and direct
address across
multiple groups
further indicates
that the partition
does not divide
the repertoire
into mutually
exclusive topics.

The resulting
interpretations
should therefore
be understood
as exploratory
descriptions
of selected
central and
network-prominent
songs, rather
than comprehensive
semantic
classifications
of the corpus.
Full community
profiles, script
counts, and
descriptions
of all 21
communities
are provided
in the
Supplementary
Information.
Table~\ref{tab:main_communities}
presents six
selected communities
to illustrate
the range of
observed
patterns and
interpretive
confidence.

\begin{table}[t]
\centering
\caption{Selected evidence-linked descriptions of six operational communities. Category labels abbreviate theme-driven, mixed content and language/form, primarily language/form, and heterogeneous/unclear; they summarize bounded AI-assisted review, not validated semantic labels or estimates of community purity. Script strata are character-composition proxies.}
\label{tab:main_communities}
\footnotesize
\setlength{\tabcolsep}{3pt}
\renewcommand{\arraystretch}{1.12}
\begin{tabular}{@{}l r >{\raggedright\arraybackslash}p{1.35cm} >{\raggedright\arraybackslash}p{2.35cm} >{\raggedright\arraybackslash}p{3.25cm} >{\raggedright\arraybackslash}p{2.1cm} >{\raggedright\arraybackslash}p{1.6cm}@{}}
\toprule
Group & $n$ & Category & Descriptive label & Content evidence & Script profile & Confidence \\
\midrule
C0 & 38 & Unclear & Unavailable love and repeated versions & Longing for unavailable love; several versions recur. & 100\% Hangul; 0\% mixed & Low \\
C2 & 243 & Theme-led & Seasonal imagery and memory & Weather/season terms with separation and memory. & 72\% Hangul; 26\% mixed & Medium \\
C6 & 350 & Mixed & Reciprocal address and relationship change & Devotion and changing relationships with reciprocal address. & 94\% Hangul; 5\% mixed & Medium \\
C8 & 372 & Theme-led & Breakup grief and place-linked memory & Separation, tears, and places tied to past relationships. & 84\% Hangul; 15\% mixed & Medium \\
C15 & 529 & Form-led & Dance, performance, and hooks & Movement, audience address, and repeated performance hooks. & 8\% Hangul; 83\% mixed & High \\
C16 & 555 & Unclear & Everyday relationship talk & Common relationship vocabulary; examples vary. & 62\% Hangul; 37\% mixed & Low \\
\bottomrule
\end{tabular}
\end{table}

\section{Boundary-spanning lyrics}
\label{sec:boundary_results}

The recurrence of related lyrical motifs across different communities
raises a further question about the organisation of the similarity
network. Although community detection provides a useful description
of regions with relatively concentrated connections, individual
songs need not be associated exclusively with their assigned
communities. Some may occupy positions that connect several
regions of the network, reflecting similarities with songs
distributed across otherwise distinguishable groups. Examining
these relationships offers an opportunity to investigate
whether boundary-spanning connectivity is associated with
greater representational novelty, or whether it instead
reflects the combination of lyrical characteristics that
are already common within the corpus.

We characterised these relationships using weighted
participation, which describes how a song's incident
edge strength is distributed across communities
\cite{GuimeraAmaral2005}. Songs with higher participation
distribute their connections more evenly across
different groups, whereas those with lower participation
concentrate a greater proportion of their connectivity
within a smaller number of communities. To examine
the characteristics of highly connected boundary
positions, we defined an operational boundary group
using the upper decile of the observed participation
distribution. The empirical threshold of
$P_i\geq0.832139$ identified 790 songs, leaving
7,101 songs in the descriptive comparison group.

The distinction between these groups was particularly
pronounced in measures describing the distribution
of connectivity. Songs in the upper participation
decile had a mean external-strength fraction of
0.762, compared with 0.483 among the remaining
songs. The difference of 0.279 had a 95\%
song-bootstrap interval of [0.274, 0.284] and
a Cliff's $\delta$ of 0.955. A similar pattern
was observed for normalized community entropy,
which averaged 0.708 in the boundary group
and 0.454 in the comparison group
(difference 0.254; interval [0.249, 0.259];
Cliff's $\delta=0.979$).

Boundary songs also connected to a greater
number of distinct external communities.
On average, they reached 11.04 communities
outside their assigned group, compared
with 6.58 for the remaining songs.
The corresponding difference was 4.46
communities (95\% song-bootstrap interval
[4.29, 4.65]; Cliff's $\delta=0.788$).
These comparisons, summarised in
Table~\ref{tab:main_boundary}, indicate
that high-participation songs occupy
positions with broadly distributed
connections rather than simply
possessing a large proportion of
external links.

However, these differences should
not be interpreted as independent
evidence that the boundary group
represents a distinct class of
songs. Participation, external-strength
fraction, community entropy, and
external reach are all derived
from the same underlying network
connections. Their relationships
are therefore partly determined
by their mathematical definitions,
and the pronounced group contrasts
are consistent with the criterion
used to select the upper
participation decile. The
comparisons provide a detailed
description of boundary
connectivity, but they do not
independently validate the
boundary classification.
Complete comparison statistics
are reported in Supplementary
Table~S9.

A more informative distinction
emerges when participation
is examined alongside measures
of embedding-relative novelty.
If songs connecting multiple
communities were also
consistently unusual within
the representation space,
we would expect higher
participation to accompany
greater distance from
community centres or
lower similarity to
immediate neighbours.
The observed relationships
did not follow this pattern.

Across the corpus,
participation showed weak
negative associations
with both centroid-relative
novelty (Spearman's
$\rho=-0.112$) and local
novelty ($\rho=-0.216$).
The upper participation
decile also exhibited
slightly lower mean
novelty scores than
the remaining songs.
Mean centroid novelty
was 0.188 in the
boundary group and
0.200 in the
comparison group,
corresponding to
a difference of
$-0.0123$ (95\%
song-bootstrap interval
[$-0.0199$, $-0.0047$]).
For local novelty,
the corresponding
means were 0.226
and 0.244, with
a difference of
$-0.0182$ (interval
[$-0.0232$, $-0.0122$]).

These findings suggest
that broad
cross-community
connectivity does
not necessarily
coincide with an
unusual position
in the embedding
space. Indeed,
within the selected
representation,
songs with higher
participation tended
to have slightly
lower novelty
scores. One
possible
interpretation
is that some
boundary-spanning
songs combine
familiar lyrical
characteristics
that are represented
in several
communities,
allowing them
to connect
different regions
without being
particularly
dissimilar
to their
neighbours.
This explanation
is consistent
with the
observed network
geometry, but
the analysis
does not establish
which lyrical
features produce
the pattern
or whether
listeners would
perceive these
songs as more
or less original.

The novelty
comparisons also
require caution
because both
measures are
defined relative
to the same
embedding
representation
used to construct
the network.
Centroid novelty
describes a
song's distance
from its nearest
community
centroid,
whereas local
novelty is
calculated from
the mean weight
of its incident
edges, including
any retained
self-loop.
Neither measure
provides an
independent
assessment
of semantic
innovation
or artistic
originality.
Furthermore,
the bootstrap
intervals
do not account
for dependencies
introduced by
shared artists,
alternate versions,
or graph
connectivity.
The corresponding
correlation
statistics
are provided
in Supplementary
Table~S10.

\begin{table}[t]
\centering
\caption{Boundary-group comparison at the q90 participation threshold (Supplementary Table~S9). Boundary songs ($n = 790$) have participation $\geq 0.832139$; all other songs ($n = 7{,}101$) form the comparison group. Mean difference is boundary minus other. Intervals are song-level bootstrap 95\% intervals for the mean difference, not for Cliff's $\delta$. The comparison is descriptive and conditional on the similarity graph and partition; the three connectivity measures are related summaries of the same graph connections and are not independent validation.}
\label{tab:main_boundary}
\footnotesize
\setlength{\tabcolsep}{3pt}
\renewcommand{\arraystretch}{1.1}
\begin{tabular}{@{}l c c c c c@{}}
\toprule
Metric & Boundary mean & Other mean & Mean difference & 95\% CI (difference) & Cliff's $\delta$ \\
\midrule
External-strength fraction & 0.7621 & 0.4832 & 0.2790 & [0.2738, 0.2838] & 0.955 \\
Community entropy & 0.7085 & 0.4545 & 0.2540 & [0.2489, 0.2586] & 0.979 \\
External communities reached & 11.0418 & 6.5801 & 4.4617 & [4.2860, 4.6511] & 0.788 \\
Centroid novelty & 0.1880 & 0.2002 & $-0.0123$ & [$-0.0199$, $-0.0047$] & $-0.098$ \\
Local novelty & 0.2258 & 0.2440 & $-0.0182$ & [$-0.0232$, $-0.0122$] & $-0.156$ \\
\bottomrule
\end{tabular}
\end{table}

Individual examples
help illustrate
how distributed
connectivity
appears within
the network.
The song
\emph{퍼플 레인
(Purple Rain)},
assigned to
C0, has a
participation
coefficient
of 0.906
and an
external-strength
fraction
of 0.923.
Its strongest
cross-community
connections
extend towards
groups described
through
night-sky
longing,
parting
and vocal
hooks,
and persistent
attachment.
These
relationships
suggest that
the song
occupies
a position
between
regions
sharing
related
emotional
and expressive
characteristics.

A second
example,
\emph{못난 내 사랑},
belongs to
C17 and
has a
participation
coefficient
of 0.898.
Its strongest
external
connections
lead towards
communities
characterised
by relationship
hurt,
difficulty
letting go,
and breakup
grief.
Although
the two
songs occupy
different
communities,
their
cross-community
connections
involve
overlapping
relationship-oriented
motifs.
This
illustrates
how the
network can
retain
relationships
between
songs
whose
computational
assignments
place them
in different
groups.

These examples
are intended
to clarify
the meaning
of boundary
connectivity
rather than
to demonstrate
a distinct
form of
lyrical
hybridity.
Both songs
were selected
from
high-participation
records,
not drawn
from a
random
sample,
and their
network
positions
do not
independently
establish
the semantic
content
of their
lyrics.
The
connections
are
consistent
with
relationships
among
recurring
motifs,
but the
interpretation
remains
conditional
on the
selected
embedding
representation
and community
partition.
Further
details
of the
selected
examples
appear
in
Supplementary
Table~S12.

Figure~\ref{fig:boundary_architecture}
illustrates
the
cross-community
connections
of these
two songs
alongside
the
group-level
effect
sizes.
The
visualisation
separates
the
individual
network
examples
from the
statistical
comparisons
between
the
upper-decile
group
and the
remaining
songs.

\begin{figure}[t]
\centering
\includegraphics[
    width=\linewidth,
    height=0.5\textheight,
    keepaspectratio
]{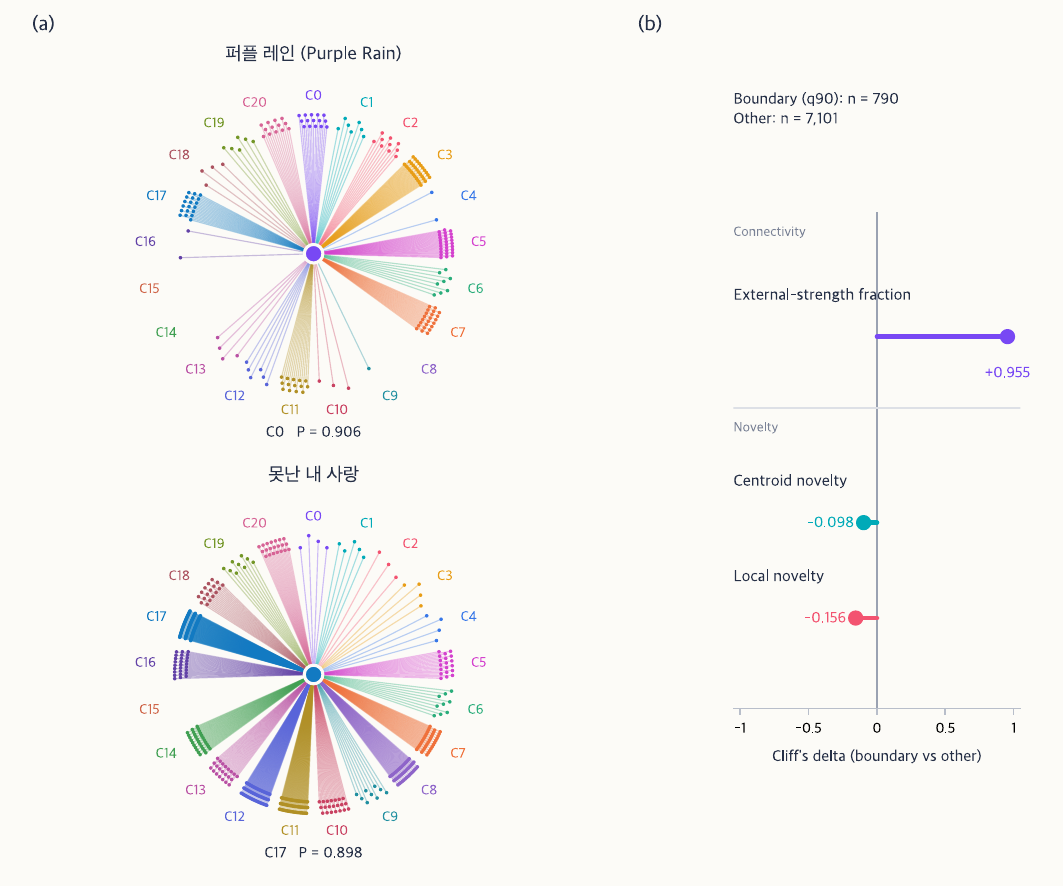}
\caption{Boundary-spanning architecture. (a) Two authentic high-participation
examples: \emph{Purple Rain} (C0; $P=0.906$) and \emph{못난 내 사랑} (C17;
$P=0.898$). Connections are coloured by the neighbouring community, edge widths
are uniform, and radial positions are diagrammatic. These are two
high-participation examples, not a high-versus-low comparison. (b) Cliff's
$\delta$ for the boundary upper-decile group ($n=790$) versus all other songs
($n=7{,}101$) for external-strength fraction ($+0.955$), centroid novelty
($-0.098$) and local novelty ($-0.156$). Panel (b) describes group-level
comparisons and does not show novelty measurements for the two songs in panel
(a). No confidence intervals are shown for Cliff's $\delta$.}
\label{fig:boundary_architecture}
\end{figure}

The distribution
of boundary
participation
can also
be examined
at the
artist
level,
although
these
summaries
are
necessarily
sensitive
to the
composition
of the
available
catalogue.
As an
illustration,
the exact
artist
string
for BTS
(\texttt{방탄소년단})
identifies
81 songs,
with a
mean
participation
of 0.657
and a
song-bootstrap
interval
of
[0.617, 0.695].
These songs
are represented
across
19 of
the
21 operational
communities.

BLACKPINK,
by comparison,
has 18
songs
in the
catalogue,
with a
mean
participation
of 0.702
and an
interval
of
[0.643, 0.756].
Its songs
appear
in nine
communities.
Although
BLACKPINK
has a
higher
mean
participation
within
these
recorded
subsets,
the
bootstrap
intervals
overlap,
and no
inferential
comparison
between
the
artists
was
performed.
Moreover,
the
number
of
communities
represented
by an
artist
is
influenced
by the
number
of
available
songs
and the
resolution
of the
selected
partition.

These
artist-level
observations
should
therefore
be
interpreted
as
descriptive
illustrations
of
catalogue
coverage
and
network
position,
not as
evidence
that
one
artist
produces
more
diverse
or
innovative
lyrics
than
another.
The
analysis
uses
exact
artist
strings
without
consolidating
aliases,
individual
members,
solo
releases,
or
collaborations,
and
the
available
catalogue
does
not
constitute
a
balanced
sample
of
artist
output.
The
corresponding
artist
summaries
are
reported
in
Supplementary
Table~S11.

Overall,
the
boundary
analysis
shows
that
the
selected
lyric
network
contains
songs
whose
similarity
relationships
extend
across
multiple
operational
communities.
Their
positions
are
associated
with
broadly
distributed
connectivity,
but
not
with
greater
embedding-relative
novelty.
This
distinction
suggests
that
cross-community
connections
may
reflect
shared
or
recombined
expressive
characteristics
rather
than
unusual
lyrical
content.
The
findings
therefore
complement
the
community-level
interpretations
by
showing
that
relationships
among
songs
cannot
be
fully
described
through
their
individual
community
assignments.

\section{Discussion}
\label{sec:discussion}

\subsection{From representation to network organization}

The findings of this study demonstrate that the computational
organization of a lyrical repertoire cannot be considered
independently of the representation through which that repertoire
is examined. Although the four embedding strategies were derived
from the same collection of songs and evaluated using a common
nearest-neighbour construction, they produced substantially
different patterns of similarity and community organization.
These differences are important because the relationships
identified within an embedding space become the foundation
for subsequent interpretations of the corpus. The choice of
representation is therefore not simply a technical preparation
step, but a methodological decision that shapes which aspects
of lyrical similarity become visible.

Nevertheless, the representation comparison does not support
a simple hierarchy in which one strategy is superior under
every criterion. Chunked-document embeddings approached
the separation achieved by whole-song truncation and
produced higher mean adjusted Rand agreement between
community-detection runs. Unique-line averaging achieved
the highest mean normalized mutual information, while
the line-based representations yielded higher modularity.
These differences reflect the fact that the evaluation
criteria describe distinct properties of the computational
structure. Separation measures the contrast between
neighbourhoods and a random-pair baseline, whereas
partition agreement describes the repeatability of
community assignments under different optimization
initializations. Modularity, in turn, characterizes the
concentration of weighted connections within a particular
partition. None of these quantities independently
establishes that the underlying representation captures
lyrical meaning more faithfully.

The selection of whole-song truncated embeddings should
therefore be understood as a reasoned choice for the
present analysis rather than a general recommendation
for computational lyric studies. Its relatively strong
neighbourhood separation and competitive partition
agreement provided a useful basis for examining the
structure of the catalogue, but the advantages of this
representation must be considered alongside its
limitations. In particular, truncation can exclude
material appearing later in a lyric, potentially
underrepresenting narrative developments, recurring
motifs, or changes in expression across the song.
The comparison consequently illustrates a broader
methodological principle: the suitability of a
textual representation depends on the analytical
question being addressed and on which properties
of the resulting relationships are considered
most informative.

The multiscale analysis reinforces this
qualification. Although the selected
24-community Markov Stability partition
showed substantial agreement with the
Louvain results, the Markov-time scan
did not identify a qualifying stability
plateau near the 21-community solution.
The agreement between methods therefore
indicates that they recover overlapping
aspects of the network's organization,
rather than independently confirming
a particular number of semantic groups.
The observed structure remains sensitive
to analytical scale, and alternative
resolutions may emphasize different
relationships within the same catalogue.

This scale dependence is not necessarily
a weakness of the network approach.
A cultural repertoire may contain
several overlapping forms of organization,
with broad expressive tendencies
coexisting alongside more localized
similarities in wording, imagery,
and lyrical form. A single partition
necessarily simplifies those
relationships. Its scientific
usefulness lies in providing
a clearly defined perspective
from which they can be examined,
not in resolving every possible
interpretation into one definitive
classification. In this sense,
the selected community structure
serves as an analytical framework
for investigating the corpus,
while the multiscale results
make the conditional nature
of that framework explicit.

\subsection{Content, script, and the meaning of communities}

The community-level findings reveal how computationally
identified groups can provide an entry point for
interpreting recurring patterns of lyrical expression.
Across the selected partition, lexical profiles and
representative songs suggest differences in imagery,
relationship narratives, forms of address, and
performance-oriented conventions. These patterns
give the network an interpretable dimension,
connecting abstract similarity relationships
with recognizable characteristics of the
underlying texts.

However, the relationship between community
membership and lyrical meaning is neither
uniform nor exclusive. Several communities
contain songs concerned with affection,
separation, longing, and remembrance,
but these broad concerns are expressed
through different combinations of
vocabulary and narrative setting.
C2, for example, brings together
seasonal and weather-related imagery
with experiences of separation and
memory, whereas the reviewed material
from C8 more prominently associates
relationship loss with remembered
places and personal experiences.
The distinction is therefore not
that one community contains longing
while another does not. Rather,
related emotional concerns appear
through different expressive patterns
that the selected representation
places in distinguishable regions
of the network.

This observation has implications
for the interpretation of
unsupervised communities in
cultural text collections.
A community label can provide
a useful summary of prominent
characteristics, but it
necessarily leaves out the
variation among individual
members and the similarities
shared with other groups.
Treating such labels as
mutually exclusive themes
would impose a degree of
semantic separation that
the available evidence
does not establish.
The communities are more
appropriately understood
as regions of relative
similarity whose contents
may exhibit recurring,
but overlapping,
expressive tendencies.

Character-script composition
adds another dimension to
this interpretation.
The observed association
between script categories
and community membership
indicates that the
distribution of Hangul,
Latin, and other
alphabetic characters
varies across the
network. This result
is relevant because
the computational
representation of
multilingual lyrics
may reflect not only
their semantic content,
but also their
linguistic and
formal characteristics.
The association does
not establish the
relative contribution
of these influences,
nor does it identify
the languages used
in individual songs.
It nevertheless
shows that textual
composition is an
important characteristic
to consider when
interpreting the
selected communities.

C6 and C15 illustrate
different ways in
which content and
textual form may
coincide. C6 combines
relationship-oriented
expressions with
a strongly
Hangul-dominant
profile, while
C15 exhibits
substantial mixed
Hangul--Latin
composition alongside
performance-oriented
vocabulary, repeated
calls, and direct
audience address.
In the latter case,
formal characteristics
such as repetition
and code-switching
may be closely
connected to the
expressive function
of the lyrics.
It would therefore
be misleading
to treat linguistic
form and lyrical
meaning as entirely
separate explanations
of the observed
community structure.

At the same time,
the uneven coherence
of the community
descriptions demonstrates
the limits of
interpretation from
lexical signatures
and selected examples.
The small C0 community
contains several
versions of closely
related material,
which may strengthen
its apparent internal
similarity without
indicating a broadly
shared expressive
category. Conversely,
the larger C16
community contains
recognizable
relationship vocabulary
but encompasses
a range of narrative
situations that
resist a single
high-confidence
interpretation.
These cases show
that concentrated
network relationships
do not guarantee
a coherent thematic
identity.

The interpretive
value of the
community analysis
therefore lies
partly in its
ability to expose
both recurring
patterns and
their exceptions.
Where lexical
evidence and
representative
lyrics support
a recognizable
description,
the community
provides a
useful perspective
on the organization
of the catalogue.
Where the
evidence is
heterogeneous,
the uncertainty
itself is
informative,
indicating that
the computational
grouping should
not be reduced
to a single
semantic label.

These interpretations
remain exploratory.
The AI-assisted
review considered
selected terms
and representative
songs rather than
a formally coded,
independently
validated sample
of the corpus.
It therefore
cannot establish
the prevalence
of particular
motifs within
each community
or determine
whether the
proposed descriptions
would be reproduced
by independent
human readers.
Such validation
would be necessary
before treating
the community
profiles as
reliable semantic
classifications.
For the present
study, they
serve instead
as evidence-informed
descriptions
that help
connect the
network structure
with observable
features of
the lyrics.

\subsection{Boundary participation and novelty}

The boundary analysis provides a complementary
perspective on the organization of the catalogue
by examining relationships that extend across
community assignments. Community detection
emphasizes regions in which connections
are relatively concentrated, but this
representation can obscure the extent
to which individual songs resemble
material distributed across several
groups. Weighted participation addresses
this aspect of the network by describing
how each song's connection strength
is distributed among communities.

The upper participation decile
exhibited broader cross-community
connectivity than the remaining
songs, as reflected in its
external-strength fraction,
community entropy, and
external reach. These findings
are consistent with the
definition of participation:
songs with more evenly
distributed connections
will often have a greater
proportion of their
similarity relationships
outside their assigned
community and reach
a wider range of groups.
The agreement among
these measures therefore
provides a detailed
structural description,
rather than independent
confirmation that the
songs possess a distinct
semantic property.

The more consequential
finding concerns the
relationship between
boundary participation
and embedding-relative
novelty. The weak
negative associations
observed for both
centroid-relative
and local novelty
indicate that songs
with more widely
distributed community
connections were not
systematically more
unusual within the
selected embedding
space. Indeed,
higher participation
was associated
with slightly lower
novelty scores.
Although these
relationships were
modest, they challenge
the assumption
that occupying
a boundary between
computational
communities necessarily
implies an
unconventional
position in the
underlying representation.

One possible
explanation is
that boundary-spanning
songs combine
characteristics
that are already
familiar in
different regions
of the catalogue.
A lyric may,
for example,
share relational
language with
one community
while also
exhibiting
imagery or
expressive
conventions
associated
with another.
Such a
combination
could produce
connections
across several
groups without
placing the
song far
from established
patterns in
the embedding
space.
The selected
song examples
are consistent
with this
possibility,
as their
external
connections
involve
communities
characterized
by overlapping
relationship-oriented
motifs.

This interpretation
remains a
hypothesis
about the
observed
network geometry,
not a
demonstrated
mechanism.
The analysis
does not
identify
which textual
features
produce
individual
cross-community
connections,
nor does
it establish
how listeners
would perceive
the originality
or meaning
of the
corresponding
songs.
Furthermore,
participation
and novelty
are derived
from the
same
representation
and graph,
making their
relationship
conditional
on the
underlying
computational
construction.

The distinction
between
connectivity
and novelty
is nevertheless
conceptually
useful.
Boundary
participation
describes
the distribution
of a song's
relationships
across an
operational
partition,
whereas
embedding-relative
novelty
describes
its position
with respect
to community
centres
or immediate
neighbours.
These
properties
need not
vary together.
Recognizing
their difference
helps avoid
interpreting
cross-community
connectivity
as a direct
measure of
lyrical
originality,
cultural
innovation,
or artistic
influence.

Taken together,
the boundary
results extend
the interpretation
of the
community
structure
without
requiring
the communities
to be treated
as isolated
semantic
categories.
They show
that songs
can maintain
substantial
similarity
relationships
across
different
regions
of the
network,
and that
these
relationships
capture
a property
distinct
from
embedding-relative
novelty.
This
perspective
is particularly
relevant
to cultural
texts,
where recurring
motifs and
expressive
conventions
may be
shared
across
otherwise
distinguishable
groups.

\subsection{Implications for computational cultural analysis}

The broader contribution of this study lies in connecting
decisions about textual representation with the
interpretation of network structure. These stages are
often considered separately: embeddings provide
numerical descriptions of documents, similarity
graphs organize the resulting relationships,
and community-detection methods summarize
their structure. Yet the present findings
show that the interpretation of the final
communities depends on decisions made
throughout this analytical sequence.

A representation that produces strongly
separated neighbourhoods need not yield
the highest modularity or the most
repeatable community assignments.
Similarly, a reproducible partition
need not correspond to a uniquely
persistent organization across
analytical scales. Even when
communities exhibit recognizable
lexical characteristics, those
patterns may reflect overlapping
combinations of content, script
composition, repetition, and
expressive form. The resulting
network is therefore best
understood as a structured
description of computational
relationships rather than
a direct map of independently
established cultural meanings.

This perspective does not
diminish the usefulness
of computational analysis.
On the contrary, it
clarifies where such
methods can make
a substantive contribution.
Embedding comparisons
make the consequences
of representation choices
visible. Network analysis
provides a means of
examining relationships
across thousands of
songs. Community profiles
help identify recurring
characteristics that
would be difficult
to survey manually,
while boundary measures
reveal connections
that extend beyond
individual group
assignments. Together,
these approaches
provide complementary
ways of examining
the organization
of a large
lyrical repertoire.

Their value,
however, depends
on maintaining
a distinction
between computational
evidence and
semantic interpretation.
Similarity in
an embedding
space is an
observable
property of
the selected
representation;
the cultural
or expressive
meaning of
that similarity
requires
additional
interpretation
and, where
stronger claims
are intended,
independent
validation.
Likewise,
community
membership
and boundary
participation
describe
relationships
within a
constructed
network,
not intrinsic
attributes
of songs
that necessarily
persist
across
representations
or corpora.

The framework
developed here
can inform
the study
of other
cultural
text collections,
particularly
those containing
repetition,
multilingual
material,
and overlapping
forms of
expression.
Its application
elsewhere
would nevertheless
require
renewed
assessment
of the
source
collection,
textual
representation,
graph
construction,
analytical
scale,
and
semantic
interpretation.
The present
results
provide
a corpus-specific
demonstration
of these
relationships,
rather
than
evidence
that the
same
community
structure
or representation
preferences
would
generalize
unchanged
to other
musical
traditions
or periods.

\subsection{Limitations}

Several limitations qualify the interpretation
and reproducibility of the findings.
The first concerns the relationship
between computational structure and
lyrical meaning. Community descriptions
were developed through an AI-assisted
exploratory examination of selected
discriminative terms and representative
songs. The procedure did not include
independent human annotation,
a formal motif-coding protocol,
or a sampling design intended
to estimate the prevalence
of particular themes.
Consequently, the resulting
descriptions cannot be treated
as validated semantic
classifications, and the
apparent coherence of
individual communities
may not extend to all
their constituent songs.

A second group
of limitations
concerns the
historical
preparation
of the corpus
and embeddings.
Although the
final analytical
collection
contains
7,891 songs,
the complete
relationship
between the
documented
catalogue-screening
procedure
and the
final exclusions
has not
been independently
verified.
The precise
truncation
limit,
tokenizer
settings,
truncation
direction,
encoding-time
normalization,
chunking
configuration,
and model
checkpoint
revision
are also
undocumented.
These
details
matter
because
different
encoding
configurations
can change
the resulting
embeddings
and, in turn,
the network
relationships
examined
in the
study.

Whole-song
truncation
introduces
a further
substantive
limitation.
When a lyric
exceeds
the encoder's
effective
input length,
material
appearing
later
in the
song
may not
contribute
to its
representation.
This
may be
particularly
consequential
for songs
whose
narrative
or expressive
characteristics
change
across
verses,
or whose
later
sections
introduce
distinctive
imagery.
The
observed
advantage
of the
whole-song
representation
on the
selected
separation
measure
should
therefore
not
be
interpreted
as
evidence
that
truncation
preserves
the
complete
semantic
content
of
a lyric.

The
historical
normalization
order
used
to
construct
community
centroids
also
remains
unresolved.
In addition,
although
the
available
software
documentation
identifies
Python 3.10.19
and
PyGenStability
0.2.5,
their
connection
to
the
environment
that
generated
the
reported
multiscale
results
has
not
been
verified.
The
available
analytical
scripts
support
inspection
of
several
reported
procedures,
but
the
original
corpus-preparation
and
embedding-generation
pipelines
have
not
been
fully
reconstructed.
Together
with
restrictions
on
redistributing
the
underlying
lyrics
and
analytical
inputs,
these
gaps
limit
complete
end-to-end
reproducibility.

A third
limitation
concerns
the
scope
of
the
corpus.
The
collection
is
derived
from
a
particular
chart
source
and
covers
a
defined
historical
period
within
the
Korean
popular-music
market.
It
should
not
be
regarded
as
a
representative
sample
of
all
K-pop
releases
or
Korean
popular
music
more
broadly.
Chart
selection,
artist
representation,
and
the
presence
of
alternate
versions
may
affect
the
distribution
of
lyrical
characteristics
and
the
structure
of
the
resulting
network.
Accordingly,
the
findings
describe
the
available
catalogue
rather
than
population-level
properties
of
the
genre.

Finally,
the
statistical
summaries
are
conditional
on
the
selected
embeddings,
graph,
and
community
partition.
Songs
associated
with
the
same
artist,
closely
related
recordings,
and
connected
regions
of
the
network
are
not
necessarily
independent
observations.
The
song-level
permutation
tests
and
bootstrap
intervals
do
not
fully
model
these
dependencies
and
should
therefore
be
interpreted
as
descriptive
rather
than
population-level
inferential
evidence.
Boundary
participation
and
novelty
are
also
sensitive
to
the
chosen
representation,
network
construction,
and
partition,
while
the
upper-decile
boundary
threshold
is
specific
to
the
observed
distribution.

The
finite
Markov-time
grid
and
number
of
optimization
trials
further
restrict
conclusions
about
community
persistence
at
untested
scales.
In
addition,
the
placement
of
new
songs
within
the
selected
representation
and
community
structure
has
not
been
validated
out
of
sample.
Future
work
could
address
these
limitations
through
more
complete
pipeline
reconstruction,
independent
semantic
annotation,
alternative
sampling
strategies,
and
validation
of
network
relationships
across
representations,
periods,
and
corpora.

\section{Conclusion}

This study examined how alternative representations
of song lyrics influence the computational
organization of a large K-pop repertoire.
Across 7,891 songs, the four representation
strategies produced different patterns
of neighbourhood similarity and
community structure. Whole-song
truncated embeddings achieved
the strongest observed separation
between neighbouring and randomly
paired songs, although other
representations performed better
on selected measures of
partition agreement or modularity.
These differences demonstrate
that the structure recovered
from a cultural corpus
depends partly on how
its textual material
is represented.

Using the selected
whole-song representation,
we examined a weighted
similarity network
through an operational
21-community partition.
The resulting groups
exhibited overlapping
patterns of lyrical
content, expressive
form, and character-script
composition, while
the multiscale analysis
showed that their
precise number
and boundaries
should not
be treated
as intrinsic
properties
of the corpus.
The boundary
analysis further
demonstrated
that songs
connecting several
communities
were not
necessarily
more unusual
within the
embedding space,
distinguishing
cross-community
connectivity
from
representation-relative
novelty.

The principal
contribution
is therefore
not a definitive
classification
of K-pop themes,
but an integrated
account of how
representation
choices,
network organization,
and interpretive
assumptions
shape the
computational
study of
lyrical relationships.
By making
these dependencies
explicit,
the framework
provides
a basis
for exploring
large cultural
text collections
without equating
computational
similarity
with independently
validated meaning.
Further work
incorporating
human semantic
assessment,
more complete
reproducibility,
and broader
corpus sampling
will be
necessary
to determine
how far
these findings
extend beyond
the present
collection.

\section*{Ethics approval and consent to participate}

Not applicable. This study uses publicly available song lyrics and does not
involve human participants, personal data, or identifiable individual
information.

\section*{Consent for publication}

Not applicable.

\section*{Data Availability}

The study used lyric records associated with Melon's Monthly Chart Ranking
Top 100. EX3exp's public
\textit{Kpop-lyric-datasets} repository
(\url{https://github.com/EX3exp/Kpop-lyric-datasets}) is identified in the
source documentation as an upstream source associated with these records.
The inspected repository snapshot is commit
\texttt{a7d141f5431fffb61c52c8ec1e48553975fa6350}.
This upstream repository is distinct from the final 7,891-song analytical
corpus. The exact relationship between the inspected snapshot and the study
input, and the complete historical preparation lineage to the analytical
corpus, have not been independently verified.

Original copyrighted lyric texts and restricted analytical inputs are not
redistributed by the authors. No claim is made that Melon distributed
the archive.

\section*{Code Availability}

Code for representation comparison, multiscale community analysis,
community characterization, and boundary/novelty summaries is available
under the MIT licence in release v1.0.0 at
\url{https://github.com/oktaykarakus/kpop-semantic-communities/releases/tag/v1.0.0}.

The package contains eight analytical scripts and excludes the study corpus,
lyric texts, embeddings, graph and partition arrays, and other restricted
analytical inputs. Execution requires compatible project inputs and
historical dependencies described in the repository documentation.
The historical corpus-preparation and embedding-generation pipelines have
not been independently reconstructed; the package therefore does not
provide complete end-to-end reproducibility.

\section*{Competing interests}

Neither author has competing interests.

\section*{Funding}

This research received no external funding.

\section*{Authors' contributions}

Oktay Karaku\c{s} conceived the study, designed the methodology, implemented
the computational pipeline, conducted the analysis, interpreted the results,
and wrote the manuscript. Can Ey\"{u}po\u{g}lu contributed to conceptualization,
formal analysis, and writing--review and editing, including work on the final
manuscript draft.

\section*{Acknowledgements}

Not applicable.

\bibliographystyle{naturemag}
\bibliography{references}

\end{document}